\documentclass[aps,prl,floatfix,twocolumn,superscriptaddress]{revtex4-1}
\pdfoutput=1
\usepackage{amsmath}
\usepackage{amssymb}
\usepackage{amstext}
\usepackage{amsopn}
\usepackage{amsfonts}
\usepackage{amsxtra}
\usepackage{color}
\usepackage{dcolumn}
\usepackage{graphicx}
\usepackage{hyperref}
\usepackage{bm}
\usepackage{pifont}
\begin{document}
%
% Title
%
\title{Do organic and other exotic superconductors fail universal scaling 
relations?}
%
% Author list
%
\author{S. V. Dordevic}
\email{Correspondence and requests for materials should be addressed
  to S.V.D. (dsasa@uakron.edu)}
\affiliation{Department of Physics, The University of Akron, Akron, Ohio 
44325, USA}
\author{D. N. Basov}
\affiliation{Department of Physics, University of California, San Diego, La 
Jolla, California 92093, USA}
\author{C. C. Homes}
\affiliation{Condensed Matter Physics and Materials Science Department,
  Brookhaven National Laboratory, Upton, New York 11973, USA}%
%
%\date{\today}
%
% The abstract goes here
%
\begin{abstract}
Universal scaling relations are of tremendous importance in science, as
they reveal fundamental laws of nature. Several such scaling relations have
recently been proposed for superconductors; however, they are not really
universal in the sense that some important families of superconductors
appear to fail the scaling relations, or obey the scaling with different
scaling pre-factors.
In particular, a large group of materials called organic (or molecular)
superconductors are a notable example. Here, we show that such apparent
violations are largely due to the fact that the required experimental
parameters were collected on different samples, with different experimental
techniques. When experimental data is taken on the same sample, using a
single experimental technique, organic superconductors, as well as all
other studied superconductors, do in fact follow universal scaling
relations.
\end{abstract}
%

%\pacs{ }%
\maketitle

\section*{Introduction}
In spite of microscopic differences, all superconductors (SC) have
one macroscopic property in common: they all conduct electricity
without resistance. Therefore, it is not unreasonable to expect
manifestations of universal behavior. We show here that when
consistent experimental parameters are used, taken on the same
sample, with a single experimental technique, {\em all} superconductors
for which the data exists, indeed follow universal scaling
relations\cite{dordevic02,homes04}.

\section*{Results}
Our scaling plots shown in Figs. 1 and 2 currently include:
elemental SC (such as Nb and Pb), cuprates (both along and
perpendicular to the CuO$_2$ planes), iron-based SC (both along
and perpendicular to iron-arsenic or iron-chalcogenide planes),
organic SC \{such as quasi-two-dimensional (BEDT-TTF)$_2$Cu(NCS)$_2$
and (BEDT-TTF)$_2$Cu[N(CN)$_2$]Br\} materials, alkali-doped fullerenes
(such as K$_3$C$_{60}$ and Rb$_3$C$_{60}$), heavy-fermion SC CeCoIn$_5$,
MgB$_2$, TiN, copper-free oxide SC Ba$_{1-x}$K$_x$BiO$_3$, negative-U
induced SC in Tl$_x$Pb$_{1-x}$Te, Y$_2$C$_2$I$_2$, etc.  Further
measurements on different SC families, both conventional and unconventional,
will serve as the ultimate test as to whether or not these scaling
relations are truly universal in nature.
%
% Section on Sr2RuO4
%
[We note in passing that the only superconductor that significantly and
systematically deviates from the scaling relations is the {\em p}-wave
superconductor Sr$_2$RuO$_4$. At this moment it is not clear whether this
violation is real, or it is due to material and/or experimental issues. It
was shown\cite{homes05} that superconductors in the clean limit do in fact
fall to the right of the scaling line, and that might be the case with
Sr$_2$RuO$_4$. However, we also note that the microwave surface impedance (MW
SI) spectra of Sr$_2$RuO$_4$ were quite unusual\cite{ormeno06,baker09}, and
to extract the penetration depth the authors had to modify the commonly-used
fitting procedure. It remains to be seen if this modification also affected
the absolute values of penetration depth ($\lambda_s$).]

Soon after superconductivity in the cuprates was discovered, Uemura
{\em et al.}\cite{uemura89} proposed the first scaling law that
related {\em ab}-plane superfluid density (or stiffness) $\rho_s$ to
superconducting critical temperature $T_c$ as $\rho_s \propto T_c$.
This scaling works for underdoped cuprates, but fails for overdoped
samples\cite{basov05rmp}. Other deviations in the cuprates were also
reported\cite{basov05rmp}. Moreover, the scaling is not followed by
other families of superconductors. Basov {\em et al.},\cite{basov94}
on the other hand, studied interplane ({\em c}-axis) response of the
cuprates and showed that the zero-temperature {\em c}-axis effective
penetration depth $\lambda_s$ is related to the {\em c}-axis DC
conductivity just above $T_c$, $\sigma_{dc}$, as
\begin{equation}
  \lambda_s \propto \sigma_{dc}^{-0.5}.
  \label{eq:basov}%
\end{equation}
This relation has been shown to be valid in a number of cuprate families.

Dordevic {\em et al.}\cite{dordevic02} extended this scaling
relation [Eq.~(\ref{eq:basov})] to other families of {\em layered} SC.
What was found based on existing experimental data was that, similar
to the cuprates, other layered SC followed similar scaling law,
albeit with a different prefactor (Fig.~2 in Ref.~\cite{dordevic02}).
The prefactor was argued to be related to the energy scale from which
the SC condensate was collected; in the cuprates the condensate was
collected from an energy range two orders of magnitude broader than
in other families \cite{basov99}. Alternatively, Schneider
interpreted the observed scaling as due to quantum criticality
\cite{schneider02}.

%
% Fig. 1 Basov plot
%
\begin{figure*}
\vspace*{-0.5cm}
\centerline{\includegraphics[width=6.5in]{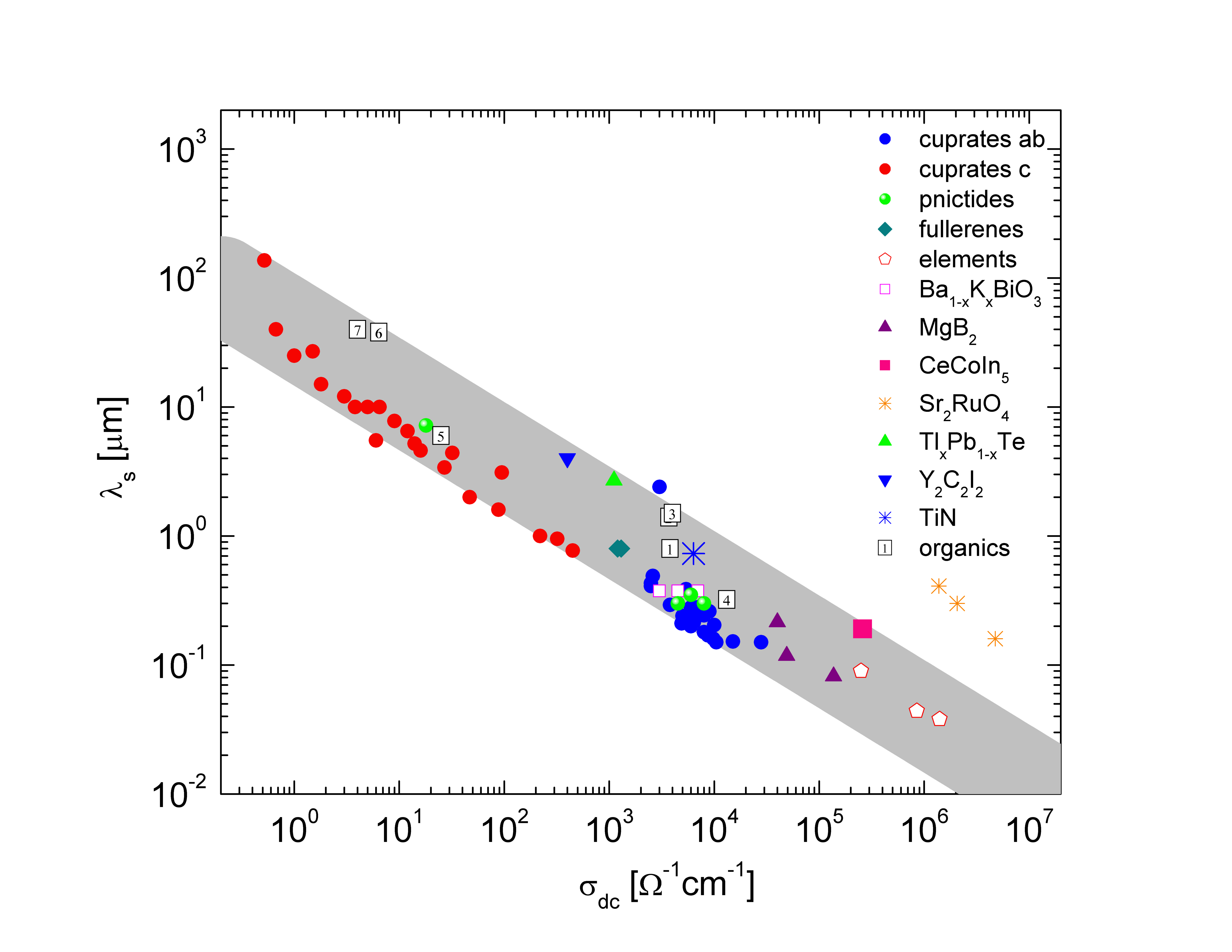}}%
\vspace*{-0.5cm}
\caption{\boldmath {\bf Basov scaling plot, Eq.~(\ref{eq:basov}).}
\unboldmath
The gray stripe corresponds to $\lambda_s = (45 \pm
25)\,\sigma_{dc}^{-0.5}$.
Only data points obtained from optical spectroscopies (IR and MW SI)
are included in the plot. The data points are from: cuprates {\em ab}-plane
\cite{homes04,pimenov99,homes12}, cuprates {\em c}-axis
\cite{homes04}, pnictides \cite{homes09,wu10}, elements
\cite{homes04}, TiN \cite{pracht12}, Ba$_{1-x}$K$_x$BiO$_3$
\cite{puchkov96}, MgB$_2$ \cite{tu01,jin03}, organic SC
\cite{dressel94,drichko02,milbradt12}, fullerenes \cite{degiorgi92},
heavy fermion CeCoIn$_5$ \cite{ormeno02}, negative-U induced SC
Tl$_x$Pb$_{1-x}$Te \cite{barker10} and Y$_2$C$_2$I$_2$
\cite{room02}.
}
\end{figure*}

Homes {\em et al.}\cite{homes04} proposed a modification to the
scaling given by Eq.~(\ref{eq:basov}), to include the SC critical
temperature $T_c$,
\begin{equation}
  \rho_s = \frac{c^2}{\lambda_s^2} \propto T_c\, \sigma_{dc},
  \label{eq:homes}%
\end{equation}
where {\it c} is the speed of light. What was found was that all
cuprate SC for which the data existed followed the scaling.
Surprisingly, both the highly conducting copper-oxygen ({\em ab}) planes
and nearly insulating out-of-plane ({\em c} axis) properties followed
the same universal scaling line. Moreover, several
elemental SC, such as Nb and Pb, also followed the same scaling
(Fig.~2 in Ref.~\cite{homes04}). More recently, iron-based SC were
also shown to follow the same scaling \cite{wu10,homes12}.

However, the so-called organic (or molecular) superconductors failed
to provide a convincing data set for the scaling Eq.~(\ref{eq:homes}),
and were not included in the original plot (Fig.~2 in
Ref.~\cite{homes04}). Several other families of superconductors,
such as dichalcogenides and heavy fermions, were also not considered
for the same reason. It has been argued that organic SC in their most
conducting planes follow different scaling laws \cite{powell04,pratt05},
such as $T_c \propto \lambda_s^{-3}$.

%
% Table I
%
\begin{center}
\begin{table*}
\caption{Parameters of organic (molecular) superconductors used in
Figs.~1 and 2. The labels in the first column correspond to the labels
in Figs.~1 and 2. The data shown is for quasi-2D organic SC, both
along their most conducting planes (1--5) and perpendicular to them
(6 and 7). Also shown are the parameters for new data points in Figs.~1 and
2. }
\vspace*{0.5cm}%
\label{organics}
\begin{tabular}{| c | l | c | c | c | c | c |}
  \hline
  % after \\: \hline or \cline{col1-col2} \cline{col3-col4} ...
  label   & organic SC & $\sigma_{dc}$ [$\Omega^{-1}cm^{-1}$] & $\lambda_s$
  [$\mu$m] & $T_c$ [K] & technique & Reference  \\
  \hline
  \hline
  1 & (BEDT-TTF)$_2$Cu(NCS)$_2$ & 3800 & 0.8 & 8.6 & MW SI (35 GHz) &
  \cite{dressel94} \\
  2 & (BEDT-TTF)$_2$Cu(NCS)$_2$ & 3700 & 1.4 & 8.3 & MW SI (60 GHz) &
  \cite{dressel94} \\
  3 & (BEDT-TTF)$_2$Cu[N(CN)$_2$]Br & 4000 & 1.5 & 11.3 & MW SI &
  \cite{dressel94} \\
  4 & (BEDT-TTF)$_2$Cu[N(CN)$_2$]Br & 13150 & 0.322 & 11 & MW SI &
  \cite{milbradt12} \\
  5 & (BEDT-TTF)$_2$I$_3$ & 25 & 6 & 8 & IR & \cite{drichko02} \\
%  \hline
  6 & (BEDT-TTF)$_2$Cu[N(CN)$_2$]Br & 6.4 & 38 & 11.3 & MW SI &
  \cite{dressel94} \\
  7 & (BEDT-TTF)$_2$Cu(NCS)$_2$ & 4 & 40 & 8.3 & MW SI & \cite{dressel94} \\
  \hline
   & K$_3$C$_{60}$   & 1200 & 0.8 & 19 & IR & \cite{degiorgi92} \\
   & Rb$_3$C$_{60}$  & 1300 & 0.8 & 29 & IR & \cite{degiorgi92} \\
  \hline
   & CeCoIn$_5$  & 260000 & 0.19 & 2.2 & MW SI & \cite{ormeno02} \\
  \hline
   & MgB$_2$  & 40000  & 0.215 & 39.6 & IR & \cite{tu01} \\
   & MgB$_2$  & 137000 & 0.082 & 39 & MW SI & \cite{jin03} \\
   & MgB$_2$  & 49000  & 0.118 & 39 & MW SI & \cite{jin03} \\
  \hline
   & Tl$_x$Pb$_{1-x}$Te  & 1111 & 2.7 & 1.4 & MW SI & \cite{barker10} \\
  \hline
   & Sr$_2$RuO$_4$  &  4760000 & 0.16 & 1.47 & MW SI & \cite{ormeno06} \\
   & Sr$_2$RuO$_4$  &  2060000 & 0.3  & 1.24 & MW SI & \cite{baker09} \\
   & Sr$_2$RuO$_4$  &  1380000 & 0.41 & 0.74 & MW SI & \cite{baker09} \\
  \hline
   & Y$_2$C$_2$I$_2$  & 400 & 4 & 10 & IR & \cite{room02} \\
  \hline
   & TiN & 6350 & 0.73 & 3.4 & IR & \cite{pracht12} \\
  \hline
  \hline
\end{tabular}
\end{table*}
\end{center}

Below we show that these discrepancies stem mostly from the fact that
the required experimental data for Eqs.~(\ref{eq:basov}) and
(\ref{eq:homes}), namely $T_c$, $\sigma_{dc}$ and $\lambda_s$, were
collected on different samples, and more importantly, using different
experimental techniques. This introduced significant scatter in data
points, and gave the impression that some families of SC did not
follow the scaling relations. The superconducting transition
temperature $T_c$ is extracted from either DC resistivity or
magnetization measurements and its values are fairly reliable and
accurate. On the other hand, the experimental values of $\sigma_{dc}$
and $\lambda_s$ can be quite problematic. The values of DC
conductivity at the transition $\sigma_{dc}$ and the zero-temperature
penetration depth $\lambda_s$ (or alternatively the superfluid
density $\rho_s$) can be extracted from a variety of experimental
techniques, and in many cases those values are significantly
different from each other. These problems seem to be most pronounced
in highly-anisotropic SC, such as the cuprates and organic SC.

The DC conductivity at the transition is most directly obtained from
transport (resistivity) measurements, but it can also be obtained
from infrared (IR) and MW SI measurements, in the $\omega \rightarrow 0$
limit. The values obtained from these spectroscopic techniques are in
some cases significantly different from the ones obtained from transport
measurements. For example, for the organic compound (TMTSF)$_2$PF$_6$
along the most conducting {\em a} axis Dressel {\em et al.} report
values obtained from both transport and IR measurements (Table~I in
Ref.~\cite{dressel97}). The value obtained from the IR measurements
is $\sigma_1(\omega\rightarrow 0) \simeq 1,100$~$\Omega^{-1}{\rm cm}^{-1}$,
whereas the DC value of conductivity is $\sigma_{dc} \simeq
80,000$~$\Omega^{-1}{\rm cm}^{-1}$, i.e. it is more than 72 times higher.
This is an extreme example, but the values for other compounds also
show large discrepancies (Table~I in Ref.~\cite{dressel97}).
Especially challenging are the IR measurements on systems with very small
and very large conductivities, and one expects large error bars
associated with them.

%
% Fig. 2 Homes plot
%
\begin{figure*}
\vspace*{-0.5cm}
\centerline{\includegraphics[width=6.5in]{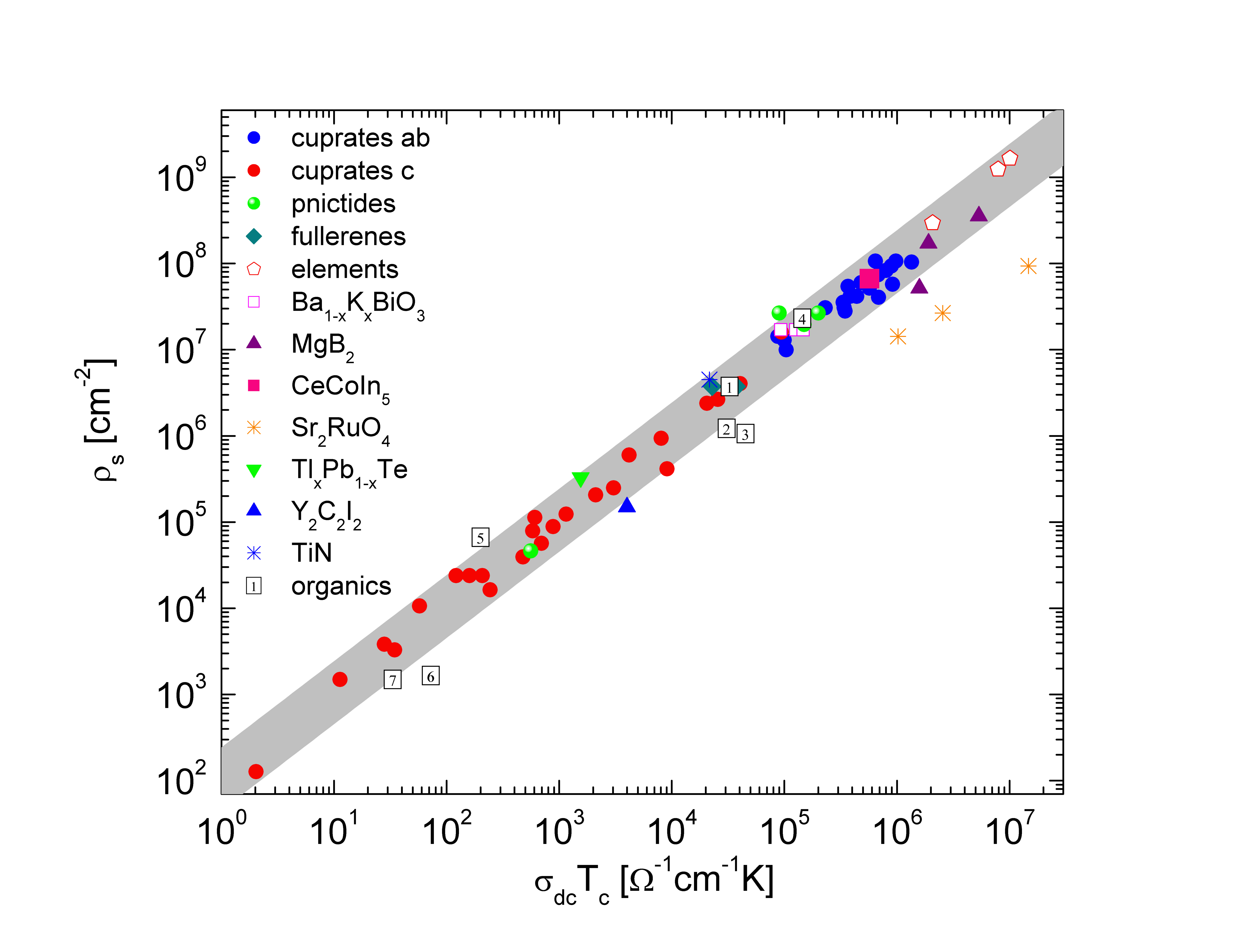}}%
\vspace*{-0.5cm}
\caption{\boldmath {\bf Homes' scaling plot, Eq.~(\ref{eq:homes}).}
\unboldmath
The gray stripe corresponds to $\rho_s = (110 \pm 60)\, T_c\, \sigma_{dc}$.
The data points are the same as in Fig.~1.}
\end{figure*}

Similar problems occur with the superfluid density. This quantity can
be extracted from optical spectroscopies (IR and MW SI), as well as
muon spin resonance ($\mu$SR) measurements. The superfluid density in
layered systems along their least conducting direction is usually
very small, which is also challenging for IR spectroscopy. Similar to
$\sigma_{dc}$, the values of $\lambda_s$ obtained from different
experimental techniques can differ significantly. For example, the
values for underdoped La$_{2-x}$Sr$_x$CuO$_4$ reported by
Panagopoulos {\em et al.} \cite{panagopoulos03} obtained using
$\mu$SR are several times smaller that those reported by IR
spectroscopy. For for the $x=0.08$ sample the $\mu$SR  value of the
penetration depth is 9.2~$\mu$m, whereas the value obtained using IR
on the sample with nominally the same doping level is 24.2~$\mu$m
(Table~I in Ref~\cite{dordevic05}).  In this case the IR
penetration depth is 2.6 times smaller, which results in superfluid
density which is almost 7 times larger [Eq.~(\ref{eq:homes})]. Similar
discrepancies are seen in other samples characterized by large
anisotropy.

The above examples illustrate the need for consistent data sets, i.e.
data obtained on the same sample, with a single experimental
technique. Therefore, in our current plots we include only such data
points. The only two experimental techniques that can deliver both
$\sigma_{dc}$ and $\lambda_s$ simultaneously are IR and MW SI.
Whenever possible, we used the data from IR spectroscopy, although in
some cases, especially for systems with low $T_c$, as well as
systems with very low and very large conductivities, we were forced to
use the MW SI data.

In Fig.~1 we re-plot the scaling from Eq.~(\ref{eq:basov}), but we now keep
only the data points taken on the same sample, with a single experimental
technique. The gray stripe shown in the picture corresponds to the $\lambda_s
= (45 \pm 25)\, \sigma_{dc}^{-0.5}$. The plot includes a variety of different
SC families, including the data for several organic SC. The values of
parameters used for new data points are shown in Table~\ref{organics}.

The scaling relation Eq.~\ref{eq:homes} is shown in Fig.~2, using the same
data from Fig.~1.  The gray stripe corresponds to  $\rho_s = (110 \pm 60)\,
T_c\, \sigma_{dc}$. A cursory inspection of the plot indicated that some
organic SC points are slightly off the scaling line (the case of
Sr$_2$RuO$_4$ was discussed above). However we do not see any {\em systematic
deviations} from the scaling, as the points are located both below and above
the scaling line. We suspect that these discrepancies are due to sample
imperfections, as well as experimental issues. For example, the data points
denoted 1 and 2 were taken on the same (BEDT-TTF)$_2$Cu(NCS)$_2$ sample, in
the same study \cite{dressel94}, at two different measurement frequencies (35
and 60~GHz, respectively); point 1 is on the scaling line, whereas point 2 is
slightly below.
Data points 3 and 4, on the other hand, have been taken on the same
compound by two different groups\cite{dressel94,milbradt12} and
point 4 (the more recent measurement) is on the scaling line, whereas
point 3 is slightly below.

\section*{Discussion}
Possible theoretical explanation of the observed scaling is a work
in progress, but some existing proposals are worth mentioning.
Tallon {\em et al.} argued that the scaling can be explained
using a dirty limit picture in which the energy gap scales
with T$_c$ \cite{tallon06,homes05}. However, it is well known that many
superconductors on the scaling plot are not in the dirty limit.
In fact, many of them are in the clean limit, and some of them
have even shown quantum oscillations. This issue of "dirtiness"
in superconductors has been discussed before\cite{basov11}.
Zaanen\cite{zaanen04} argued that the superconducting transition
temperature in cuprates is high because the normal state in these
systems is as viscous as is allowed by the laws of quantum mechanics.
Zaanen also introduced the notion of Plankian dissipation in the
cuprates\cite{zaanen04}. However, this proposal does not explain why
{\em all} superconductors, not just the curpates, follow the same
scaling. Imry {\em et al.} demonstrated that the scaling may
be recovered in an inhomogeneous superconductor in the limit of
small intergrain resistance in a simple granular superconductor
model\cite{imry12}.  The scaling relation Eq.~(\ref{eq:homes})
has also been derived using the gauge/gravity duality for a holographic
superconductor\cite{erdmenger12}.

In summary, we have shown that when consistent data sets are used,
{\em all} superconductors for which the data sets exist do indeed
follow universal scaling relations that span more than seven orders
of magnitude. Future experiments on other (exotic) SC will serve as
important test of validity of scaling relations, and will verify if
they are truly universal.

\section*{Methods}
Data points shown in Figs. 1 and 2 are collected from different literature
sources, either IR or MW SI measurements. Those two experimental techniques
can simultaneously deliver the two parameters needed for scaling
Eq.~(\ref{eq:homes}), namely the optical conductivity at T$_c$, $\sigma_{dc}
\equiv \sigma_1(\omega \rightarrow 0)$, and the superfluid density $\rho_s$
(or the penetration depth $\lambda_s$). This selection assures that the
required parameters were collected on the same sample, in a single
measurement, without the use of contacts.

\section*{acknowledgments}
The authors thank C. Petrovic for pointing out the heavy fermion
data. S.V.D. acknowledges the support from The University of Akron FRG.
Research supported by the U.S. Department of Energy, Office of Basic
Energy Sciences, Division of Materials Sciences and Engineering under
Contract No. DE-AC02-98CH10886.
D.N.B. acknowledges support from the National Science Foundation (NSF 
1005493).

%%%%%%%%%%%%%%%%%%%%%%%%%%%%%%%%%%%%%%%%%%%%%%%%%%%%%%%%%%%%%%%%%
%
% The bibliography (BibTeX)
%
%\bibliography{organics}

%merlin.mbs apsrev4-1.bst 2010-07-25 4.21a (PWD, AO, DPC) hacked
%Control: key (0)
%Control: author (8) initials jnrlst
%Control: editor formatted (1) identically to author
%Control: production of article title (-1) disabled
%Control: page (0) single
%Control: year (1) truncated
%Control: production of eprint (0) enabled
\providecommand{\noopsort}[1]{}\providecommand{\singleletter}[1]{#1}

\end{document}